\documentclass{ifacconf}
\usepackage{subfig}
\usepackage{amssymb}
\usepackage{mathrsfs}
\usepackage{graphicx}      
\usepackage[hyphens]{url}
\usepackage{algorithm}
\usepackage{colortbl}
\usepackage{algpseudocode}
\usepackage{amsmath}
\usepackage{natbib}        

\newtheorem{assumption}{Assumption}
\newtheorem{definition}{Definition}
\newtheorem{remark}{Remark}

\begin{document}
\begin{frontmatter}

\title{Data-Driven Decentralized Control Design for Discrete-Time Large-Scale Systems\thanksref{footnoteinfo}} 

\thanks[footnoteinfo]{This research was partially supported by the National Science Foundation, under NSF CNS Award no. 2223035,  NSF CAREER Award no. 2143351, and NSF IIS Award no. 2331938.}

\author[First]{Jiaping Liao} 
\author[Second]{Shuaizheng Lu} 
\author[First]{Tao Wang}
\author[Second]{Weiming Xiang}

\address[First]{School of Electrical Engineering, Southwest Jiaotong University, China. (e-mails: \{liaojiaping,taowang\}@swjtu.edu.cn) }
\address[Second]{School of Computer and Cyber Sciences, Augusta University, Augusta GA 30912, USA. (e-mails: \{shulu,wxiang\}@augusta.edu)}

\begin{abstract}                
In this paper, a data-driven approach is developed for controller design for a class of discrete-time large-scale systems, where a large-scale system can be expressed in an equivalent data-driven form and the decentralized controllers can be parameterized by the data collected from its subsystems, i.e., system state, control input, and interconnection input. Based on the developed data-driven method and the Lyapunov approach, a data-driven semi-definite programming problem is constructed to obtain decentralized stabilizing controllers. The proposed approach has been validated on a mass-spring chain model, with the significant advantage of avoiding extensive modeling processes.
\end{abstract}


\end{frontmatter}

\section{Introduction}

The large-scale control system is a class of complex systems that are comprised of numerous interacting subsystems that can be found in extensive engineering applications such as robotics \cite{kordestani2021recent}, power systems \cite{rajan2021primary},  and so on. On the other hand, data-driven model identification is a process of using data analysis and machine learning methods to estimate the unknown parameters of a mathematical model based on available data, which can be used for system prediction, control, and optimization. Previous research on data-driven methods has mainly focused on studying the identification and control of individual dynamical systems such as in \cite{Y9}. Given the rapid development of large-scale systems in various fields, this work aims to develop data-driven control for large-scale systems. 

Data-driven methods have been widely used in dynamical system identification. For instance, subspace system identification is a system identification method that uses matrix decomposition or singular value decomposition (SVD) to reduce the dimensionality of high-dimensional data, extract subspace information, and construct mathematical models. It has been widely applied in signal processing, control, and other fields. In \cite{Y3}, it provides a comprehensive overview of the current state of subspace identification methods in open-loop and closed-loop systems, with a detailed exposition of the standard model and commonly used mathematical methods in subspace identification. In \cite{Y4}, it focuses on closed-loop subspace identification methods and provides a clear overview of some successful methods proposed in the past decade. In addition, the authors trace the common origins of these methods, highlight their differences, and compare them based on simulation examples and real data. In subspace system identification, a crucial question is determining when the matrix created from a state sequence has rows that are linearly independent of the matrix formed from an input sequence and any finite number of its shifts. In \cite{Y2}, it demonstrates that if one of the response signal's components in a controllable linear time-invariant system is persistently exciting of a high enough order, then the windows of the signal cover the entire system behavior without exception. 

Despite the widespread applications of subspace model identification as a means of data-driven modeling, there recently exist promising results that design control directly using data. In \cite{Y7}, the computational issues of minimum energy control for linear systems are studied based on experimental data, and an open-loop minimum energy control input is designed. In \cite{Y8}, the result proposes a data-driven solution to the discrete infinite horizon linear quadratic regulator (LQR) problem and validates the design methodology's practicality on an uninterrupted power supply through experiments. The previous results have demonstrated that it is feasible to directly design control laws from data. In a more recent result in \cite{Y9}, the data-driven control design is systematically introduced, along with the resolution of control problems such as stabilization of state and output feedback, and linear quadratic regulation. This developed data-driven controller design method has been generalized to other systems such as switched systems \cite{rotulo2022online}, nonlinear polynomial systems\cite{guo2021data}, and controller design scenarios such as output feedback controller design \cite{hu2023data}.  Furthermore, the method exhibits robustness against measurement noise and can be applied to stabilize unstable equilibrium points of nonlinear systems. In \cite{Y11}, the author provides a more in-depth discussion of the linear quadratic regulator problem for linear systems with unknown dynamics.
In \cite{Y1,Y5,Y10}, they have made significant contributions to the study of large-scale system control, including uncertainty management, control performance optimization, and reachable set-constrained analysis. Furthermore,  \cite{Y6} also discusses the small-gain theorem for stability analysis of interconnected systems, which is associated with large-scale systems.

In this paper, we propose a data-driven decentralized controller design for discrete-time large-scale systems, where the design procedure is formulated in terms of semi-definite problems with the aid of collected system data matrices. The remainder of the paper is organized as follows:
Preliminaries are given in Section II. The main results, data-driven decentralized controller design, are presented in Section III. The application to interconnected spring-mass systems is given in Section IV. The conclusion is given in Section V.

\emph{Notation:} Let $\mathbb{R}$ denote the field of real numbers, and $\mathbb{R}^{n}$ stands for
the vector space of all $n$-tuples of real numbers and $\mathbb{R}^{n \times n}$ is the space
of $n \times n$ matrices with real entries. $\mathbb{Z}$ represents the set of 
integers. The notation $A\succ0$ means $A$
is real symmetric and positive definite.
$A\succ B$ means that $A -B \succ 0$. $A^{\top}$ denotes the transpose of $A$. $A^\dagger$ is the Moore–Penrose inverse of $A$. In symmetric block matrices, we use $\star$ as an ellipsis for the terms that are introduced by symmetry, and $\mathrm{diag}\{\cdot\}$ stands for a block-diagonal matrix.
For two integers 
$k_1$ and $k_2$, $k_1 \le k_2$, we define $\mathbb{I}[k_1, k_2] \triangleq \{k_1, k_{1+1},\ldots, k_2\}$.

\section{Preliminaries}
\subsection{System Description}
Consider a class of discrete-time large-scale systems with $M$ subsystems in the form of 
\begin{align}\label{sys_1}
    x_i(k+1) = A_ix_i(k)+B_iu_i(k)+\sum_{j = 1}^M {{G_{ij}}{g_{ij}}( {{x_j}}(k) )} , 
\end{align}
where ${x_i} \in {\mathbb{R}^{{n_i}}}$, ${u_i} \in {\mathbb{R}^{{m_i}}}$, $i\in \mathbb{I}[1,M]$ are the state vector and  control input vector of the $i$th subsystem. Matrices $\{A_i, B_i, G_{ij}\}$, $i,j \in \mathbb{I}[1,M] $, are constant matrices with appropriate dimensions which are unknown or unavailable for controller design. Nonlinearities $g_{ij}(x_j(k))$, $i \ne j$, $i,j \in \mathbb{I}[1,M] $, indicate the information exchange between subsystem $i$ and $j$ at time instant $k$ as the interconnection input. In the rest of this work, the state $x_i$, control input $u_i$, and interconnection input $g_{ij}(x_j)$ are assumed to be measurable for data-driven controller design.  

\begin{assumption}\label{assumption_1}
We assume that the interconnection input $g_{ij}(x_j(k))$, $i \ne j$, $i,j \in \mathbb{I}[1,M]$ of the large-scale system (\ref{sys_1}) satisfy the following conditions:
\begin{enumerate}
	    \item ${g_{ij}}(0) = 0$, $\forall i,j \in \mathbb{I}[1,M]$.
         \item $\left\| {{g_{ij}}\left( r \right) - {g_{ij}}\left( s \right)} \right\| \le\left\| W_{ij} {\left( {r - s} \right)} \right\|$, $\forall r,s \in {\mathbb{R}^{{n_i}}}$, $\forall i,j \in \mathbb{I}[1,M]$, where $W_{ij}$ are known real constant matrices.
\end{enumerate}
\end{assumption}

\begin{remark}
    Based on Assumption \ref{assumption_1}, the following result can be obtained immediately
    \begin{align}
		\left\| {{g_{ij}}( {{x_j}} )} \right\| \le \left\| W_{ij} {{x_j}} \right\|,~\forall {x_j} \in {\mathbb{R}^{{n_j}}} ,
	\end{align} 
for any $i,j \in \mathbb{I}[1,M]$. In data-driven design scenarios where $g_{ij}(x_j)$ in system model are unavailable, the matrices $W_{ij}$ can be estimated offline via measurable state $x_j$ and interconnection input $g_{ij}(x_j)$, e.g., $W_{ij} = \bar w_{ij}I$, where $\bar w_{ij} = \max\{\left\|g_{ij}(x_j)\right\|/\left\|x_j\right\|\}$. 
\end{remark}
	
For the sake of simplicity, we use the following notations for the rest of the paper:
\begin{align*}
    \mathcal{G}_{i} &= \begin{bmatrix}
       G_{i1} & G_{i1} & \cdots & G_{iM} 
    \end{bmatrix} ,
\\
    \phi_{i}^{\top}(k) &= \begin{bmatrix}
        g_{i1}^{\top}(x_1(k) )  &  \cdots & g_{iM}^{\top}(x_M(k) )
    \end{bmatrix} \in \mathbb{R}^{\ell_i} .
\end{align*}

As a result, system (\ref{sys_1}) can be rewritten into a more compact form of
\begin{align}\label{sys_2}
    x_i(k+1) = A_ix_i(k)+B_iu_i(k)+ \mathcal{G}_{i}\phi_{i}(k) ,
\end{align}
where $\{A_i, B_i, \mathcal{G}_{i}\}$, $i \in \mathbb{I}[1,M] $ are unknown matrices and the pair $\{A_i, [B_i,\mathcal{G}_{i}]\}$ are assumed to be controllable.

\subsection{Data Matrices}
Given a signal $s : \mathbb{Z} \to \mathbb{R}^n$, we denote  sequence $s_{[k,k+T]}$ as
\begin{align}\label{data_w}
   s_{[k,k+T]} = \{
       s(k), s(k+1) , \ldots, s(k+T)
  \}  ,
\end{align}
and matrix $S_{[k,k+T]}$ in the form of 
\begin{align}\label{data_w}
   S_{[k,k+T]} = \begin{bmatrix}
       s(k) & s(k+1) & \cdots & s(k+T)
   \end{bmatrix}  ,
\end{align}
where $k$, $T \in \mathbb{Z}$. By the data matrix in the form of (\ref{data_w}), we define the data matrices for subsystem $i \in \mathbb{I}[1,M] $ as follows
\begin{align}
    U_{i,[0,T-1]} = \begin{bmatrix}
    u_i(0) & u_i(1) & \cdots & u_i(T-1)
    \end{bmatrix} ,\label{data_u}
    \\
    \Phi_{i,[0,T-1]} = \begin{bmatrix}
    \phi_i(0) & \phi_i(1)  & \cdots & \phi_i(T-1)
    \end{bmatrix},
    \\
        X_{i,[0,T-1]} = \begin{bmatrix}
    x_i(0) &   x_i(1) &  \cdots & x_i(T-1)
    \end{bmatrix} ,
\end{align}
which are starting at $k=0$ and ending at $k=T$. Moreover, we define $X_{i,[1,T]}$ as
\begin{align}
X_{i,[1,T]} = \begin{bmatrix}
    x_i(1) &   x_i(2) &  \cdots & x_i(T)
    \end{bmatrix} .
    \label{data_x_1}
\end{align}

It is noted that system (\ref{sys_2}) leads to the following result
\begin{align}
    X_{i,[1,T]} = 
 \begin{bmatrix}
     B_{i} & \mathcal{G}_{i} & A_i
 \end{bmatrix}
 \begin{bmatrix}
     U_{i,[0,T-1]} \\ \Phi_{i,[0,T-1]} \\ X_{i,[0,T-1]} 
 \end{bmatrix} .
\end{align}

Throughout the work, the following rank condition plays a key role in decentralized data-driven controller design 
\begin{align}\label{rank}
   \mathrm{rank}  
   \left(\begin{bmatrix}
     U_{i,[0,T-1]} \\ \Phi_{i,[0,T-1]} \\ X_{i,[0,T-1]} 
 \end{bmatrix}\right)
 = n_i + \ell_i +m_i .
\end{align}

\begin{remark}
    The above rank condition (\ref{rank}) is a generalized version of the fundamental property established in \cite{Y2}. If we enforce the interconnection inputs to zero, i.e., $\phi_i(k) = 0$, $\forall k \in \mathbb{Z}$, leading to $\Phi_{i,[0,T-1]} = 0$, rank condition (\ref{rank}) is then reduced to  
    \begin{align}\
   \mathrm{rank}  
   \left(\begin{bmatrix}
     U_{i,[0,T-1]} \\ X_{i,[0,T-1]} 
 \end{bmatrix}\right)
 = n_i + m_i .
\end{align}
which is the result proposed in \cite{Y2} for LTI systems.
\end{remark} 

\begin{definition} \cite{Y2}
    The signal $s_{[0,T-1]} \in \mathbb{R}^{n}$ is persistently exciting of order $L$ if the matrix
    \begin{align}
        S_{0,T-1,L} = \begin{bmatrix}
            s(0) & s(1) & \cdots & s(T-L)
            \\
            s(1) & s(2) & \cdots & s(T-L+1)
            \\
            \vdots & \vdots & \ddots & \vdots
            \\
            s(L-1) & s(L) & \cdots & s(T-1)
        \end{bmatrix} ,
    \end{align}
    is of full rank $nL$. 
\end{definition}

The following result implies that the collected system data with sufficient persistently exciting of order can guarantee the rank condition (\ref{rank}) holds.

\begin{lem} \label{lemma_1}
Consider system (\ref{sys_2}), if the control and interconnection input pair for subsystem $i \in \mathbb{I}[1,M]$, i.e., $\{u_{i,[0,T-1]},\phi_{i,[0,T-1]}\}$, is persistently exciting of order $n_i +  1$, then the rank condition (\ref{rank}) holds.
\end{lem}
\begin{pf}
  System (\ref{sys_2}) can be rewritten into 
\begin{align}
    x_i(k+1) = A_ix_i(k) + \tilde{B}_i \tilde u_i(k) ,
\end{align}
where $\tilde B_i = [B_i, \mathcal{G}_i]$ and $\tilde u_i(k) = [u_i^{\top}(k), \phi_i^{\top}(k)]^{\top}$. Then, the result can be straightforwardly obtained by Lemma 1 in \cite{Y9}. The proof is complete.     
\end{pf}
\begin{remark}
    As elaborated in \cite{Y9}, if $T_i$ are taken sufficiently large, then the rank condition (\ref{rank}) turns out to be satisfied. In particular, the requirement is that $T_i \ge (m_i +\ell_i)(n_i+1) + n_i$, to establish the persistence of excitation condition. 
\end{remark}

\section{Main Results}

Given data matrices $U_{i,[0,T-1]}$, $\Phi_{i,[0,T-1]}$, $X_{i,[0,T-1]}$, and $X_{i,[1,T]}$, which are defined in (\ref{data_u})--(\ref{data_x_1}), our first question is if we can exactly reconstruct the system matrices $\{A_i, B_i, G_{ij}\}$, $i,j \in \mathbb{I}[1,M]$ out of these data matrices. In other words, we are looking for a solution of $\begin{bmatrix}
     B_{i}^{*} & \mathcal{G}_{i}^{*} & A_i^{*}
 \end{bmatrix}$ for the Least Square (LS) problem
\begin{align} \label{ls}
    \min_{[B_{i},  \mathcal{G}_{i},  A_i]}\left\|X_{i,[1,T]} - 
 \begin{bmatrix}
     B_{i} & \mathcal{G}_{i} & A_i
 \end{bmatrix}
 \begin{bmatrix}
     U_{i,[0,T-1]} \\ \Phi_{i,[0,T-1]} \\ X_{i,[0,T-1]} 
 \end{bmatrix} \right\| , 
\end{align}
which results in a minimum value of 0, i.e., 
\begin{align}\label{opt_0}
    \left\|X_{i,[1,T]} - 
 \begin{bmatrix}
     B_{i}^{*} & \mathcal{G}_{i}^{*}  & A_i^{*} 
 \end{bmatrix}
 \begin{bmatrix}
     U_{i,[0,T-1]} \\ \Phi_{i,[0,T-1]} \\ X_{i,[0,T-1]} 
 \end{bmatrix} \right\| = 0 . 
\end{align}

The following lemma trivially implies that we can exactly represent system (\ref{sys_1}), i.e., obtaining matrices $\begin{bmatrix}
     B_{i}^{*} & \mathcal{G}_{i}^{*} & A_i^{*}
 \end{bmatrix}$ out of data matrices $U_{i,[0,T-1]}$, $\Phi_{i,[0,T-1]}$, $X_{i,[0,T-1]}$, and $X_{i,[1,T]}$,  as long as the control and interconnection inputs are persistently exciting of sufficiently high order. 
 \begin{lem}\label{thm1}
     Given input data $\{u_{i,[0,T-1]},\phi_{i,[0,T-1]}\}$ persistently exciting of order $n_i + 1$ and measurable system state data $x_{i,[0,T]}$, then large-scale system (\ref{sys_1}) can be exactly represented in the following form of 
     \begin{align} \label{thm1_1}
    x_i(k+1) =  X_{i,[1,T]}\begin{bmatrix}
     U_{i,[0,T-1]} \\ \Phi_{i,[0,T-1]} \\ X_{i,[0,T-1]} 
 \end{bmatrix}^{\dagger} 
    \begin{bmatrix}
        u_i(k)
        \\
        \phi_i(k)
        \\
        x_i(k)
    \end{bmatrix} .
\end{align}
 \end{lem}
\begin{pf}
The proof is in Appendix \ref{appendix_A}.
\end{pf}

Lemma \ref{thm1} implies that the open-loop system (\ref{sys_1}) can be represented as a data-driven identification form of (\ref{thm1_1}) based on sufficient system state $x_i$, system input $u_i$, and interconnection input $g_{ij}(x_j)$ collected from the system. Following the same idea, a closed-loop system with decentralized controllers $u_i(k) = K_i x_i(k)$ can be equivalently represented with persistently exciting data of sufficiently high order, as shown in the following theorem. 

\begin{thm}\label{thm2}
     Given input data $\{u_{i,[0,T-1]},\phi_{i,[0,T-1]}\}$ persistently exciting of order $n_i +  1$ and measurable system state data $x_{i,[0,T]}$, then large-scale closed-loop system (\ref{sys_1}) with decentralized state feedback $u_i(k) = K_ix_i(k)$ can be exactly represented in the following form of
     \begin{align}\label{thm2_1}
         x_i(k+1) = X_{i,[1,T]}(H_{i,1}x_i(k)+ H_{i,2}\phi_{i}(k)) ,
     \end{align}
     where $H_{i,1}$ and $H_{i,2}$ are matrices satisfying 
     \begin{align} \label{thm2_2}
         \begin{bmatrix}
     U_{i,[0,T-1]} \\ \Phi_{i,[0,T-1]} \\ X_{i,[0,T-1]} 
 \end{bmatrix}H_{i,1}
 & =
          \begin{bmatrix}
        K_i \\ 0 \\ I
    \end{bmatrix} ,
 \\
 \begin{bmatrix}\label{thm2_3}
     U_{i,[0,T-1]} \\ \Phi_{i,[0,T-1]} \\ X_{i,[0,T-1]} 
 \end{bmatrix}H_{i,2}
 &=
 \begin{bmatrix}
        0 \\ I \\ 0
    \end{bmatrix}.
     \end{align}
Furthermore, the decentralized state feedback law can be represented as $u_i(k)=U_{i,[0,T-1]}H_{i,1}x_i(k)$.
\end{thm}
\begin{pf}
    Given that $\{u_{i,[0,T-1]},\phi_{i,[0,T-1]}\}$ is persistently exciting of order $n_i +1$,  Lemma \ref{lemma_1} implies that the following rank condition holds
    \begin{align}
       \mathrm{rank}\left( \begin{bmatrix}
     U_{i,[0,T-1]} \\ \Phi_{i,[0,T-1]} \\ X_{i,[0,T-1]} 
 \end{bmatrix}\right) = n_i + \ell_i + m_i .
 \end{align}
 
 Then, according to Rouch\'e–Capelli theorem, there always exist matrices $H_{i,1}$ and $H_{i,2}$ such that (\ref{thm2_2}) and (\ref{thm2_3}) hold, respectively. 

Therefore, we have
 \begin{align*}
(A_i+B_iK_i)x_i(k) &=
    \begin{bmatrix}
        B_i & \mathcal{G}_i & A_i
    \end{bmatrix}
      \begin{bmatrix}
        K_i \\ 0 \\ I
    \end{bmatrix}x_i(k) 
        \\
    &= \begin{bmatrix}
        B_i & \mathcal{G}_i & A_i 
    \end{bmatrix}
     \begin{bmatrix}
     U_{i,[0,T-1]} \\ \Phi_{i,[0,T-1]} \\ X_{i,[0,T-1]} 
 \end{bmatrix}H_{i,1}x_i(k)
 \\
& =X_{i,[1,T]}H_{i,1}x_i(k) ,
\end{align*}
and 
\begin{align*}
    \mathcal{G}_{i}\phi_i(k) & = \begin{bmatrix}
        B_i & \mathcal{G}_i & A_i
    \end{bmatrix}
    \begin{bmatrix}
        0 \\ I \\ 0
    \end{bmatrix} \phi_i(k)
    \\
    &= \begin{bmatrix}
        B_i & \mathcal{G}_i & A_i 
    \end{bmatrix}
    \begin{bmatrix}
     U_{i,[0,T-1]} \\ \Phi_{i,[0,T-1]} \\ X_{i,[0,T-1]} 
 \end{bmatrix}H_{i,2}\phi_i(k)
 \\
 & =X_{i,[1,T]}H_{i,2}\phi_i(k) .
\end{align*}

As a result, large-scale system (\ref{sys_1}) with decentralized state feedback in the form of $  x_i(k+1) = (A_i+B_iK_i)x_i(k) + \mathcal{G}_{i}\phi_i(k)$ arrives at
\begin{align}
x_i(k+1) = X_{i,[1,T]}(H_{i,1}x_i(k)+ H_{i,2}\phi_{i}(k)). 
\end{align}

Furthermore,  the decentralized state feedback gains $K_i = U_{i,[0,T-1]}H_{i,1}$ can be obtained directly from (\ref{thm2_2}). The proof is complete.
\end{pf}

Theorem \ref{thm2} provides the crucial result that the closed-loop large-scale system with decentralized controller $u_i(k)=K_ix_i(k)$ can be equivalently described in the form of (\ref{thm2_1}) using the matrices of states, system inputs, and interconnection inputs if the input data is persistently exciting of order $n_i+1$. In fact, Lemma \ref{thm1} can be used to design decentralized feedback controller gains $K_i$, $i \in \mathbb{I}[1,M]$ from data by reconstructing system matrices $\{A_i, B_i, G_{ij}\}$, $i,j \in \mathbb{I}[1,M] $ and the following model-based design procedures. From the observation of Theorem \ref{thm2}, the feedback controller gains $K_i$, $i \in \mathbb{I}[1,M]$ can be parameterized through data matrices $U_{i,[0,T-1]}$, $\Phi_{i,[0,T-1]}$, and $X_{i,[0,T-1]}$, thus the search of proper decentralized feedback controller gains $K_i$, $i \in \mathbb{I}[1,M]$ that guarantees
stability and performance specifications can be incorporated into a one-step design procedure without identifying the parametric model of the system.

Based on Theorem \ref{thm2}, the following will discuss the identification-free decentralized data-driven controller synthesis design for large-scale systems. 

\begin{thm}\label{thm3}
    Consider a large-scale system in the form of (\ref{sys_1}) with input data $\{u_{i,[0,T-1]},\phi_{i,[0,T-1]}\}$ persistently exciting of order $n_i +  1$, measurable system state data $x_{i,[0,T]}$, and $H_{i,2}$ satisfying (\ref{thm2_3}), if there exist matrices $Q_i$, $i \in \mathbb{I}[1,M]$ such that 
    \begin{align}\label{thm3_1}
\begin{bmatrix}
        -S_i  & 0 & Q_{i}^{\top}X_{i,[1,T]}^{\top} & S_i\mathcal{W}_i^{\top}      
        \\
        \star & -I & H_{i,2}^{\top}X_{i,[1,T]}^{\top} & 0
        \\
        \star & \star & -S_i & 0
        \\
        \star & \star & \star & -I
    \end{bmatrix} \prec 0 ,
 \end{align}
in which $\mathcal{W}_{i} = \begin{bmatrix}
    W_{1i}^{\top} & W_{2i}^{\top} & \cdots & W_{Mi}^{\top}
\end{bmatrix}^{\top}$ and $S_i = X_{i,[0,T-1]}Q_i$, and $Q_i$, $i \in \mathbb{I}[1,M]$ satisfy 
\begin{align}\label{thm3_2}
\Phi_{i,[0,T-1]}Q_i= 0 ,
    \end{align}
then the decentralized state feedback gains 
\begin{align}\label{thm3_3}
         K_i = U_{i,[0,T-1]}Q_iS_i^{-1} ,
\end{align}
asymptotically stabilize system (\ref{sys_1}). 
\end{thm}

\begin{pf}
    By Theorem \ref{thm2}, given the persistently exciting sequence $\{u_{i,[0,T-1]},\phi_{i,[0,T-1]}\}$  of order $n_i +  1$, the large-scale system (\ref{sys_1}) with decentralized state feedback $u_i(k) = K_ix_i(k)$ can be equivalently represented in the form of (\ref{thm2_1}) as long as (\ref{thm2_2}) and (\ref{thm2_3}) hold. 

    By letting $H_{i,1} = Q_i(X_{i,[0,T-1]}Q_i )^{-1}$ which indicates $X_{i,[0,T-1]}H_{i,1} =I$, along with (\ref{thm3_3}) implying $U_{i,[0,T-1]}H_{i,1} = K_i$ and (\ref{thm3_2}) leading to $\Phi_{i,[0,T-1]}H_{i,1} = 0$, we can establish $H_{i,1}$ satisfying (\ref{thm2_2}). Together with $H_{i,2}$, which always exists according to the proof line in Theorem \ref{thm2}, satisfying (\ref{thm2_3}), we can obtain the data-driven representation in the form of (\ref{thm2_1}) with feedback gains (\ref{thm3_3}). Next, we will prove the asymptotic stability of (\ref{thm2_1}).  

    Letting the Lyapunov functional candidate as follows 
    \begin{equation}
		V(x(k)) = \sum\limits_{i = 1}^M {{V_i}\left( {{x_i(k)}} \right)}  = \sum\limits_{i = 1}^M {x_i^{\top}}(k){S_i^{-1}}{x_i(k)} 
	\end{equation}
where $S_i^{-1} \succ 0$, $\forall i \in \mathbb{I}[1,M]$, and defining $\Delta V_i(k) = V_i(x(k+1))-V_i(x(k))$, one can obtain that
\begin{align}
    \Delta V_i(k)= \begin{bmatrix}
        x_i(k) \\ \phi_{i}(k)
    \end{bmatrix}^{\top}
    \begin{bmatrix}
       Z_{i,1} & Z_{i,2}
        \\
        \star & Z_{i,3}         
    \end{bmatrix}
    \begin{bmatrix}
        x_i(k) \\ \phi_{i}(k)
    \end{bmatrix} ,
\end{align}
where $Z_{i,1}$,$Z_{i,2}$, and $Z_{i,3}$ are defined by
\begin{align*}
    &Z_{i,1}=H_{i,1}^{\top}X_{i,[1,T]}^{\top}S_i^{-1}X_{i,[1,T]}H_{i,1} -S_i^{-1} ,
    \\
    &Z_{i,2} = H_{i,1}^{\top}X_{i,[1,T]}^{\top}S_i^{-1}X_{i,[1,T]}H_{i,2} ,
    \\
    &Z_{i,3}= H^{\top}_{i,2}X^{\top}_{i,[1,T]}S_i^{-1}X_{i,[1,T]}H_{i,2} .
\end{align*}

Further, we define an auxiliary variable as
\begin{align*}
    \Theta_i &= \sum_{j=1}^{M}x^{\top}_j(k)W_{ij}^{\top}W_{ij}x_j(k) - \phi_{i}^{\top}(k)\phi_{i}(k) ,
\end{align*}
which satisfies $\Theta_i \ge 0$ from Assumption \ref{assumption_1}.

Due to the following fact
\begin{align*}
\sum\limits_{i = 1}^M \sum_{j=1}^{M}x^{\top}_j(k)W_{ij}^{\top}W_{ij}x_j(k) = \sum\limits_{i = 1}^M \sum_{j=1}^{M}x^{\top}_i(k)W_{ji}^{\top}W_{ji}x_i(k)  ,
\end{align*}
we can obtain that
\begin{align}
     \sum_{i=1}^{M}\hat \Theta_i = \sum_{i=1}^{M} \Theta_i \ge 0 ,
\end{align}
where $\hat \Theta_i$ is defined as follows
\begin{align*}
    \hat \Theta_i = \sum_{j=1}^{M}x^{\top}_i(k)W_{ji}^{\top}W_{ji}x_i(k)  - \phi_{i}^{\top}(k)\phi_{i}(k) .
\end{align*}

Thus,  it leads to
\begin{align*}
   &\sum_{i=1}^{M}\left(\Delta V_i(k) + \hat \Theta_i \right)
    \\
    =& 
    \sum_{i=1}^{M}
    \begin{bmatrix}
        x_i(k) \\ \phi_{i}(k)
    \end{bmatrix}^{\top}
    \begin{bmatrix}
       Z_{i,1}+ \mathcal{W}_{i}^{\top}\mathcal{W}_{i} & Z_{i,2}
        \\
        \star & Z_{i,3} -I        
    \end{bmatrix}
    \begin{bmatrix}
        x_i(k) \\ \phi_{i}(k)
    \end{bmatrix} ,
\end{align*}
where $\mathcal{W}_{i} = \begin{bmatrix}
    W_{1i}^{\top} & W_{2i}^{\top} & \cdots & W_{Mi}^{\top} 
\end{bmatrix}^{\top}$.

Pre- and post-multiplying (\ref{thm3_1}) with $\mathrm{diag}
    \{S_i^{-1}, I, I, I\}$, one can obtain
\begin{align}
    \begin{bmatrix}
        -S_i^{-1}  & 0 & (Q_{i}S_i^{-1})^{\top}X_{i,[1,T]}^{\top} & \mathcal{W}_i^{\top}      
        \\
         \star & -I & H_{i,2}^{\top}X_{i,[1,T]}^{\top} & 0
        \\
        \star & \star & -S_i & 0
        \\
        \star & \star & \star & -I
    \end{bmatrix} \prec 0 .
\end{align}

Using $H_{i,1} = Q_i(X_{i,[0,T-1]}Q_i )^{-1} = Q_iS_i^{-1}$ and Schur complement formula, it leads to 
\begin{align}
    \begin{bmatrix}
       Z_{i,1}+ \mathcal{W}_{i}^{\top}\mathcal{W}_{i} & Z_{i,2}
        \\
        \star & Z_{i,3} -I        
    \end{bmatrix} \prec 0 .
\end{align}

Therefore, we have 
\begin{align}
   \Delta V(k) =  \sum_{i=1}^{M}\Delta V_{i}(k) \le \sum_{i=1}^{M}\left(\Delta V_i(k) + \hat \Theta_i \right) <0 .
\end{align}

Thus the closed-loop system represented in the equivalent data-driven form of (\ref{thm2_1}) is asymptotically stable according to the standard Lyapunov Theorem. The proof is complete.
\end{pf}
\begin{remark}
    As observed in the above theorem, the decentralized data-driven controller gains $K_i$, $i \in \mathbb{I}[1,M]$ can be obtained via solving semi-definite programming conditions formulated fully with the help of data. The key of this data-driven approach is that the data should ensure the rank condition (\ref{rank}) holds, i.e., persistently exciting of order $n_i+1$.  In terms of implementation, the data can be collected during a sufficiently long period $[0,T]$ for the open-loop system as long as the rank condition (\ref{rank}) is satisfied, e.g., under random initial states with random inputs as in \cite{Y9}. Moreover, this approach is effective in addressing control issues in large-scale systems. However, it may be relatively conservative since it is designed for general large-scale systems with each subsystem interconnected with one another. To alleviate the conservativeness, the controller design can be further optimized based on the actual interconnections between subsystems.    
\end{remark}

\section{Application to Interconnected Spring-Mass Systems}

The spring-mass model is commonly used as a dynamic system model in control theory, particularly to describe the motion of particles. It has been extensively applied in control systems and related fields.
Each mass in the spring-mass system has an independent input $u_i$, and neighboring masses are connected in sequence in the longitudinal direction by springs, corresponding to interconnection terms between subsystems. When the relative positions $s_i$ of neighboring masses are non-zero, they exert forces on each other, i.e. 
\begin{equation}
	f_i=k(s_i-s_j),~\forall i\in \mathbb{I}[1,M],~j\in \mathbb{I}[1,M] \cap \{i-1,i+1\} .
\end{equation}

Accordingly, the dynamic characteristics of the mass can be incorporated to derive the differential equations of the spring-mass system composed of $n$ masses and $n-1$ springs in the spring-mass model.
	\begin{equation}\label{Chain}
        \small
		\left\{ \begin{aligned}
			{{\dot s}_i} &= {v_i},~ i \in \mathbb{I}[1,M]\\
			{{\dot v}_1} &= \frac{{{u_1}}}{{{m_1}}} - b{v_1} - \frac{k}{m_1}\left( {{s_1} - {s_2}} \right)\\
			{{\dot v}_i} &= \frac{{{u_i}}}{{{m_i}}} - b{v_i} - \frac{k}{m_i}\left( {2{s_i} - {s_{i - 1}} - {s_{i + 1}}} \right),
			~i \in \mathbb{I}[2,M-1] \\
			{{\dot v}_M} &= \frac{{{u_M}}}{{{m_M}}} - b{v_M} - \frac{k}{m_M}\left( {{s_M} - {s_{M - 1}}} \right)
		\end{aligned} \right. 
	\end{equation}
where $m_i$ is the mass, $k$ is the elasticity coefficient, and $b$ is the aerodynamic drag coefficient. $s_i$ and $v_i$ denote the position and velocity of each mass, i.e. the state of the $i$th subsystem. The following table shows the parameters of the chain of masses in the simulation.
\begin{table}[htbp]
	\caption{Parameters of Spring-Mass System} \label{Table1}
	\centering
		\begin{center}
			\begin{tabular}{ccc}
				\cline{1-3}
				\textbf{Symbol} & \textbf{Value} & \textbf{Unit}     \\ \cline{1-3}
				$m$      & 1     & $\mathrm{kg}$   \\
				$b$      & $-0.1$  & $\mathrm{N \cdot s/(kg \cdot m)}$\\
				$k$      & 0.1   & $\mathrm{N/m}$   \\ \cline{1-3}
			\end{tabular}
		\end{center}
\end{table}

\begin{remark}
	The differential equations of the spring-mass model, as presented in (\ref{Chain}), will be constructed in MATLAB based on the parameters listed in Table \ref{Table1}. It is noteworthy that this is a continuous model, in this example, we will discretize with a sampling interval $0.01 s$. If we consider each mass as a subsystem, with interconnection terms also existing between neighboring masses due to springs, it can be described in the form of a large-scale system (\ref{sys_1}). 
	\end{remark}

As mentioned earlier, this study uses a data parametric controller approach that allows the controller to be obtained by solving the semi-definite problem in Theorem \ref{thm3}.
 In contrast to the model-based control method, this approach employs several data-driven design steps, and the following process will be carried out for controller design.
\begin{enumerate}
	\item Set the data sample interval and the input injection interval, they should be consistent, denote $T_k$.
	\item Generate a sequence of input sequences $u_i(k), k\in [0,T-1]$ that satisfy the row full rank condition. Form input matrices $U_{i,[0,T-1]},\forall i\in \mathbb{I}[1,M]$ and inject them sequentially into the spring-mass model.
	\item Sample the states $[s_i,v_i]^{\top}$, $\forall i\in \mathbb{I}[1,M]$ and interconnections $[s_j,v_j]^{\top}$, $j\in \mathbb{I}[1,M]\cap \{i-1,i+1\}$ at the moment and the next moment of each input injection, and form matrices ${U_{i,[0,T-1]}}$, ${X_{i,[0,T-1]}}$, ${\Phi_{i,[0,T-1]}}$, ${X_{i,[1,T]}}$.
	\item Verify that the rank condition (\ref{rank}), if not,  return to Step 2.
	\item Given  ${{U_{i,[0,T-1]}},{X_{i,[0,T-1]}},{\Phi_{i,[0,T-1]}},{X_{i,[1,T]}}}$ and applying Theorem \ref{thm3}, calculate the discrete state feedback controller $K_i$ of the $i$th subsystem.
	\item With closed-loop control laws $u_i=K_ie_i$, $i\in \mathbb{I}[1,M]$, the spring-mass model tracks a given velocity $v_r$, where $e_i=[s_i-s_r,v_i-v_r]^{\top}$ denotes position and velocity errors.
\end{enumerate}

Setting the sampling interval $T_k=0.01 \mathrm{s}$, the discrete state feedback controllers obtained data-based approach are shown below. In the simulation setup, we assumed that five masses are with different random initial states located in $[49,51]$, which aim to track the velocity target of $v_r=50 \mathrm{m/s}$.

\begin{table}[htb] 
	\caption{State feedback controllers for subsystems}\label{tableK}
	\centering
	\begin{tabular}{c|c}
\hline
\textbf{Sub-Controllers} & {\textbf{Data-Driven Controller Gains}} \\ \hline
$K_1 $                               & $[-409.87,~-89.86 ]$            \\
$K_2 $                               &  $[-409.94,~-89.91]   $          \\
$K_3$                                &  $[-405.00,~-89.37]$           \\
$K_4  $                              &  $[-411.28,~-92.00]$           \\
$K_5$                                &  $[-410.48,~-91.32]$            \\ \hline
	\end{tabular}
\end{table}


\begin{figure}[htp]
	\centering
\includegraphics[width=0.45\textwidth]{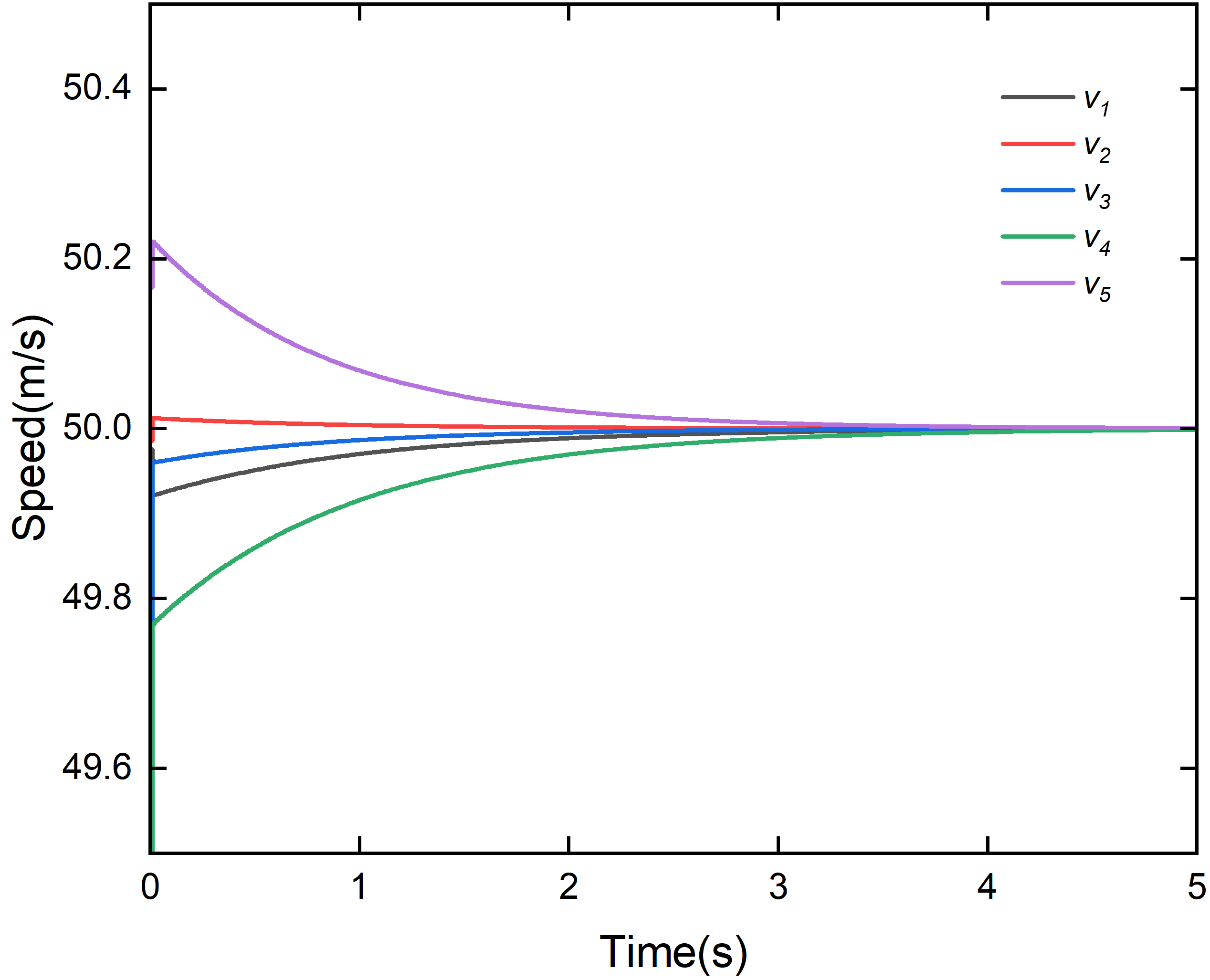} 
	\caption{State responses of five masses under decentralized state feedback controllers. The velocities of five masses converge to the target of $v_r=50 \mathrm{m/s}$ under the decentralized data-driven controller.}\label{SimulationResultDtP}
\end{figure}

Applying the data-driven controllers in Table \ref{tableK} to System (\ref{Chain}), the simulation results are presented in Fig. \ref{SimulationResultDtP}. 
In the simulation results,  the data-driven decentralized controller allows the spring-mass model to converge to the tracking speed, taking about 4 seconds, which shows the efficiency of the decentralized data-driven control strategy. 
 
\section{Conclusions}

This work developed a decentralized data-driven controller design for large-scale systems such as the spring-mass system, establishing the feasibility and stability of solving semi-definite problems directly from data collected from systems. The key lies in two steps: first, parameterizing the undetermined controllers for the subsystems based on their states, inputs, and interconnection terms, and second, constructing a semi-definite problem for the stabilization of large-scale systems using the aforementioned parameterized subsystem controllers, and solving them. The former step can serve as a standard procedure to lay the foundation for further research, whereas the latter step can be extended for different types of large-scale systems or performance objectives by establishing new semi-positive definite problems. However, the proposed method has limitations, as it does not consider noise in the data matrix, also time delays, uncertainties, and other issues between communications of subsystems are expected to be incorporated into this data-driven design framework. Hence, additional efforts will be devoted to our future study to adequately address the limitations and challenging problems of the current work. 


\appendix
\section{Proof of Lemma \ref{thm1}} 
\label{appendix_A}

We let  $\Xi = \begin{bmatrix}
     B_{i} & \mathcal{G}_{i}  & A_i
 \end{bmatrix}$ and
\begin{align*} 
    Y_{i,[0,T-1]} = \begin{bmatrix}
     U_{i,[0,T-1]} \\ \Phi_{i,[0,T-1]} \\ X_{i,[0,T-1]} 
 \end{bmatrix} .
\end{align*} 
Then, the LS problem (\ref{ls}) becomes
\begin{align}\label{pthm_1}
    \min_{\Xi_i}\left\|X_{i,[1,T]} - 
 \Xi_i
 Y_{i,[0,T-1]} \right\| .
\end{align}

In order to exactly represent system (\ref{sys_1}), we need to find a solution $\Xi_i^{*}$ with zero approximation error, i.e., $\left\|X_{i,[1,T]} - 
 \Xi_i^*
 Y_{i,[0,T-1]}\right\| =0$.

According to the results in \cite{penrose1956best}, the solution to the LS problem (\ref{pthm_1}) is
\begin{align} \label{pthm_2}
    \Xi_i^{*} = X_{i,[1,T]}Y_{i,[0,T-1]}^{\dagger} + (I-Y_{i,[0,T-1]}^{\dagger}Y_{i,[0,T-1]})w ,
\end{align}
for any vector $w \in \mathbb{R}^{T}$. 

Given that $\{u_{i,[0,T-1]},\phi_{i,[0,T-1]}\}$ is persistently exciting of order $n_i + 1$ which yields $\mathrm{rank}(Y_{i,[0,T-1]}) = n_i + \ell_i + m_i $ based on Lemma \ref{lemma_1}, we have 
\begin{align} 
    I-Y_{i,[0,T-1]}^{\dagger}Y_{i,[0,T-1]} = 0 .
\end{align} 

Therefore, (\ref{pthm_2}) implies that 
\begin{align}
    \Xi_i^{*} = X_{i,[1,T]}Y_{i,[0,T-1]}^{\dagger}  ,
\end{align}
in which case we can conclude that
\begin{align}
    \left\|X_{i,[1,T]} - 
 X_{i,[1,T]}Y_{i,[0,T-1]}^{\dagger} 
 Y_{i,[0,T-1]} \right\|  = 0 . 
\end{align} 

As a result, system (\ref{sys_2}) can be exactly expressed as
\begin{align}
    x_i(k+1) =  X_{i,[1,T]}\begin{bmatrix}
     U_{i,[0,T-1]} \\ \Phi_{i,[0,T-1]} \\ X_{i,[0,T-1]} 
 \end{bmatrix}^{\dagger} 
    \begin{bmatrix}
        u_i(k)
        \\
        \phi_i(k)
        \\
        x_i(k)
    \end{bmatrix} ,
\end{align}
with zero approximation error. The proof is complete.


\bibliography{ref}             

                                                   







\end{document}